\documentclass[twocolumn,trackchanges]{aastex631}
\usepackage{amsmath,amssymb}
\usepackage{listings,jvlisting}
\usepackage[legacycolonsymbols]{mathtools}
\usepackage{comment}
\usepackage{bm}

\received{\today}
\revised{}
\accepted{}
\shorttitle{Investigation of off-CIE plasma in a giant UX Ari flare with \textit{NICER}}
\shortauthors{Kurihara et al.}
\graphicspath{{./}{figures/}}

\begin{document}

\title{\uppercase{Investigation of non-equilibrium ionization plasma during a
giant flare of UX Arietis triggered with \textit{MAXI} and observed with \textit{NICER}}}

\correspondingauthor{M. Kurihara}
\email{kurihara@ac.jaxa.jp}

\author[0000-0002-3133-9053]{Miki Kurihara}
\affiliation{Department of Astronomy, Graduate School of Science, The University of
Tokyo,\\7-3-1 Hongo, Bunkyo-ku, Tokyo 113-0033, Japan}
\affiliation{Japan Aerospace Exploration Agency, Institute of Space and Astronautical
Science,\\3-1-1 Yoshinodai, Chuo-ku, Sagamihara, Kanagawa 252-5210, Japan}
\author[0000-0002-0207-9010]{Wataru Buz Iwakiri}
\affiliation{Chiba University, Inage-ku, Chiba, Chiba, 263-8522 Japan}
\author[0000-0002-9184-5556]{Masahiro Tsujimoto}
\affiliation{Japan Aerospace Exploration Agency, Institute of Space and Astronautical
Science,\\3-1-1 Yoshinodai, Chuo-ku, Sagamihara, Kanagawa 252-5210, Japan}
\author[0000-0002-5352-7178]{Ken Ebisawa}
\affiliation{Department of Astronomy, Graduate School of Science, The University of
Tokyo,\\7-3-1 Hongo, Bunkyo-ku, Tokyo 113-0033, Japan}
\affiliation{Japan Aerospace Exploration Agency, Institute of Space and Astronautical Science,\\
3-1-1 Yoshinodai, Chuo-ku, Sagamihara, Kanagawa 252-5210, Japan}
\author[0000-0002-1276-2403]{Shin Toriumi}
\affiliation{Japan Aerospace Exploration Agency, Institute of Space and Astronautical Science,\\
3-1-1 Yoshinodai, Chuo-ku, Sagamihara, Kanagawa 252-5210, Japan}
\author[0000-0001-7891-3916]{Shinsuke Imada}
\affiliation{Department of Earth and Planetary Science, Graduate School of Science, The
University of Tokyo,\\7-3-1 Hongo, Bunkyo-ku, Tokyo 113-0033, Japan}
\author[0000-0001-9943-0024]{Yohko Tsuboi}
\affiliation{Department of Physics, Faculty of Science and Engineering, Chuo University,
\\1-13-27 Kasuga, Bunkyo-ku, Tokyo 112-8551, Japan}
\author{Kazuki Usui}
\affiliation{Department of Astronomy, Graduate School of Science, The University of
Tokyo,\\7-3-1 Hongo, Bunkyo-ku, Tokyo 113-0033, Japan}
\affiliation{Kavli Institute for the Physics and Mathematics of the Universe,\\
5-1-5 Kashiwanoha, Kashiwa, Chiba, 277-8583, Japan}
\author[0000-0001-7115-2819]{Keith C. Gendreau}
\affiliation{Astrophysics Science Division, NASA's Goddard Space Flight
Center, \\Greenbelt, MD 20771, USA}
\author{Zaven Arzoumanian}
\affiliation{Astrophysics Science Division, NASA's Goddard Space Flight
Center, \\Greenbelt, MD 20771, USA}

\begin{abstract}
 We detected a giant X-ray flare from the RS-CVn type binary star UX Ari using
 \textit{MAXI} on 2020 August 17 and started a series of \textit{NICER} observations 89
 minutes later.  For a week, the entire duration of the flare was covered with 32
 snapshot observations including the rising phase. The X-ray luminosity reached
 2$\times$10$^{33}$~erg~s$^{-1}$ and the entire energy release was $\sim 10^{38}$~erg in
 the 0.5--8.0~keV band. X-ray spectra characterized by continuum emission with lines
 of \ion{Fe}{25} He$\alpha$ and \ion{Fe}{26} Ly$\alpha$ were obtained. We found that the
 temperature peaks before that of the flux, which suggests that the period of plasma
 formation in the magnetic flare loop was captured. Using the continuum information
 (temperature, flux, and their delay time), we estimated the flare loop size to be $\sim
 3 \times 10^{11}$~cm and the peak electron density to be $\sim
 4\times10^{10}$~cm$^{-3}$. Furthermore, using the line ratio of \ion{Fe}{25} and
 \ion{Fe}{26}, we investigated any potential indications of deviation from
 collisional ionization equilibrium (CIE). The X-ray spectra were consistent with CIE
 plasma throughout the flare, but the possibility of an ionizing plasma away from
 CIE was not rejected in the flux rising phase.
\end{abstract}

\section{Introduction} \label{s1}
Solar and stellar flares are considered to be caused by the same physical mechanisms but
with different scales (e.g.; \citealt{fletcher2011,benz2010}). The accumulated magnetic
energy is impulsively released through magnetic reconnection and plasma instability
\citep{priest2002,shibata2011}. In the standard model for the eruptive flares, known as
the CSHKP model \citep{carmichael1964,sturrock1966,hirayama1974,kopp1976}, the released
energy is partially converted into the kinetic energy of charged particles at the
reconnection site, which then bombard the flare loop top and cause an impulsive increase
in temperature. The particles move downward along the loop and collide with the dense
chromosphere. The evaporated matter fills the loop, causing the flux to peak. After
the initial rise, the temperature decreases conductively and radiatively.

Because of this dynamic process, it is expected and often observed that the temperature
peak precedes the flux peak (e.g.;
\citealt{maggio2000,pillitteri2022,stelzer2022}). This rising phase is the time
when physical constraints of flares can be obtained independently from and more
stringently than the decay phase. \citet{reale2007} proposed a model to describe the
behavior of X-ray light curves during flares and derived analytical formulae relating
observables with physical properties of flares and applied them to observations. In
X-ray observations of stellar flares, the observables are often limited to values
obtained from the continuum emission (such as temperature and volume emission measure;
hereafter called ``continuum observables''), but those from the line emission (``line
observables'') will carry independent information about the flare dynamics. However,
this has been hampered by the lack of spectral resolution or photon statistics
\citep{favata2000a,maggio2000,stelzer2002b}.

Of particular interest are line ratios that deviate from collisional ionization
equilibrium (CIE), as it is an expected consequence of the dynamic nature of flares,
thus providing new keys to understanding flare physics
\citep{imada2011b,reale2008a,bradshaw2003}. There are some observational claims
\citep{Kawate2016,imada2021} of non-equilibrium ionization (NEI) plasma in solar flares,
but such observations are limited despite the continuous and intensive observations of
the Sun. The largest obstacle is the short duration of NEI conditions expected to last
only for $\mathcal{O}(10^{2}~\mathrm{s})$.

\medskip

Stellar flare observations could be more advantageous. The electron temperature is $\sim
10$ times larger than the Sun during flares \citep{feldman1995}, which makes the
equilibrium time scales longer for the same electron density, as more charged
ions dominate the population and the ionization/recombination rate coefficient
decreases.  In order to exploit this advantage, we need X-ray observations from the
rising phase of stellar flares using a telescope with a large collecting area and the
capability to distinguish major lines of different charges. 

We break through this limitation by combining two X-ray instruments onboard the
International Space Station (ISS): the Monitor of All-sky X-ray Image (\textit{MAXI};
\citealt{Matsuoka2009}) and the Neutron star Interior Composition Explorer
(\textit{NICER}; \citealt{Gendreau2016}). \textit{MAXI} updates all-sky X-ray images
every 92 minutes, which has led to 167 detections of stellar flares from 30 objects since
2009\cite{}. \textit{NICER} has an ability to conduct fast maneuvers and collect
photons with a larger effective area and a better spectral resolution than \textit{MAXI}.

We have developed two systems, in which \textit{MAXI} detects transient events and
\textit{NICER} responds in the shortest achievable time. One is OHMAN (Orbiting
High-energy Monitor Alert Network), in which the triggering from \textit{MAXI} to
\textit{NICER} is automated within the ISS. OHMAN has been in operation since 2023 but
is effective only for very bright sources. The other is MANGA (\textit{MAXI} And
\textit{NICER} Ground Alert) since 2017, in which a human decision is inserted based on
the \textit{MAXI} data downlinked to the ground. MANGA still makes human-triggered
observations faster than any previous approaches. In our study, we use MANGA for stellar
flare observations (e.g., \citealt{ATel12248,sasaki2021}).

\medskip

We present here the result of the MANGA observation made with the shortest response time
to date for stellar flares. We detected a giant flare from UX Ari with \textit{MAXI} and
initiated \textit{NICER} observations well before the flux peak. UX Ari is an RS-CVn
type binary system consisting of a K0\,IV and G5\,V star with a 6.4~day orbital period
\citep{hummel2017}. The source is known to cause occasional flares in X-rays
\citep{franciosini2001,Tsuboi2016,gudel1999,tsuru1989}. The peak flare luminosity of the
present data reaches 2$\times$10$^{33}$~erg~s$^{-1}$ in the 0.5--8.0~keV band, which is
$\sim$10 times larger than any flares observed with X-ray pointing observations of this
source. The goal of this study is to examine the plasma development throughout the flare
and investigate any hints of the off-CIE plasma using the X-ray spectra. Throughout the
paper, we use \texttt{AtomDB} \citep{smith2001,foster2017} for plasma calculations. The
quoted uncertainties are for 90\% statistical error.

\section{Observations} \label{s2}
\subsection{Instrument} \label{s2-1}
\textit{NICER} is a payload onboard the ISS equipped with an X-ray Timing Instrument
(XTI). XTI is an assembly of 56 sets of X-ray concentrators \citep{Okajima2016} and
silicon drift detectors \citep{Prigozhin2016}, of which 52 are in routine use. The
energy range is 0.2--12.0 keV. The effective area and the energy resolution in FWHM at
6~keV are $\sim$ 600 cm$^2$ and 137 eV, respectively.

\textit{NICER} is useful for wide astrophysical applications. In particular, the
operational flexibility, the large effective area, and the fast detector readout allow
spectroscopic observations of eruptive events uncompromised by limitations of photon
statistics and detector dynamic range. These features make \textit{NICER} particularly
well suited for stellar flare observations
\citep{namekata2020,sasaki2021,hamaguchi2023}.

\subsection{Observation and data reduction} \label{s2-2}
We captured a flux increase of UX Ari based on the \textit{MAXI} Nova-Alert system
\citep{negoro2016} recorded on 2020 August 17 11:51:54 (UT). Using the MANGA system, the
first \textit{NICER} observation was executed 89 minutes later. \textit{NICER} continued
observations for a week (ObsIDs: 3100380101--3100380106) until the X-ray count rate
returned to the pre-flare level. \textit{NICER} visited
the source 32 times, which we refer to as snapshot 0 to 31. The on-source exposure
ranged from 101 to 1286~s with an average of 752~s and the photon counts ranged
from $2.2 \times 10^{4}$ to $2.0 \times 10^{6}$ with an average of $4.0 \times 10^{5}$
counts. Thanks to the large effective area of \textit{NICER}, it can accumulate
sufficient counts even with short exposure times.

We applied level 2 processing using the \texttt{nicerl2} tasks to incorporate the latest
calibration and event screening and produced level 3 products using the \texttt{nicerl3}
tasks in \texttt{HEASoft} version 6.32.1 \citep{2014ascl.soft08004N}. The level
3 products include the telescope and detector response files, binned source spectrum,
and background spectrum based on the \texttt{SCORPEON} model, which we generated and
used for each snapshot.

\section{Analysis} \label{s3}
\subsection{Light curve and spectra} \label{s3-1}
\begin{figure}[!hbtp]
\includegraphics[width=1.0\columnwidth,clip]{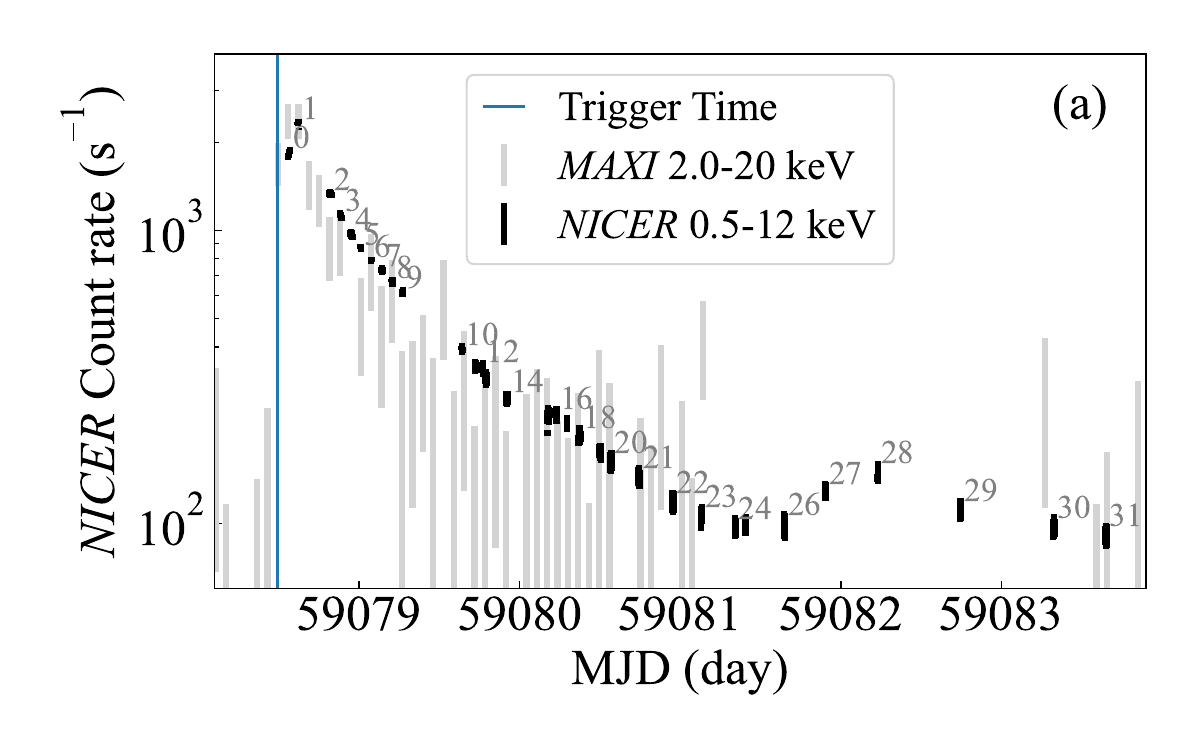}
\includegraphics[width=1.0\columnwidth,clip]{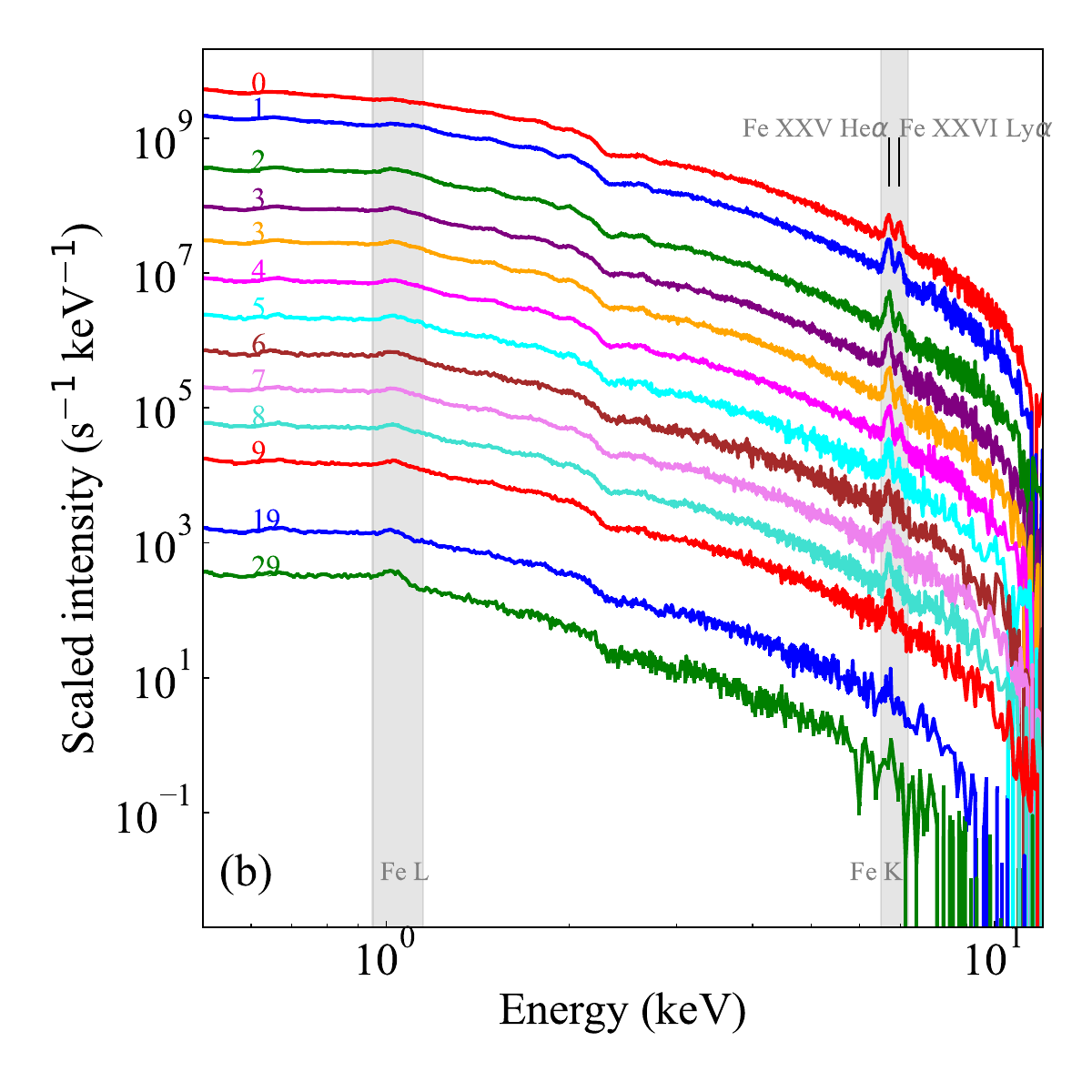}
 \caption{(a) \textit{NICER} (black) and \textit{MAXI} (grey) light
 curves. \textit{MAXI} count rate is scaled to match with \textit{NICER}. The labels
 0--31 are given for each \textit{NICER} snapshot. 
 (b) \textit{NICER} X-ray spectra with different colors for different snapshots. The
 intensity of each spectrum is offset incrementally by 0.5 dex for clarity.}
\label{f01}
\end{figure}

Fig.~\ref{f01} (a) shows the X-ray light curve of the data set. Snapshot 0 captures the
rising part of the flare, 1 captures the flux peak, and the rest shows the decaying
part. There is a slight re-brightening during 14--16 and 26--29.  The peak luminosity
and the e-folding time of the \textit{NICER} light curve are $2 \times 10^{33}$ erg
s$^{-1}$ and $2.4 \times 10^{4}$ s, respectively, deriving the total released energy in
0.5--12 keV to be $\sim 10^{38}$ erg.

Fig.~\ref{f01} (b) shows the X-ray spectra for the first ten snapshots (0--9) and two
selected later ones(19 and 29). The spectra are characterized by continuum emission
with lines. The most conspicuous feature is found at the Fe K band, in which
\ion{Fe}{25} He$\alpha$ and \ion{Fe}{26} Ly$\alpha$ emission lines are identified at 6.7
and 7.0~keV.

The development of the line pair traces the decreasing plasma temperature. The intensity
ratio of \ion{Fe}{26} Ly$\alpha$ against \ion{Fe}{25} He$\alpha$ monotonically decreases
with time, suggesting that the Fe$^{25+}$ (H-like) population is taken over by Fe$^{24+}$
(He-like) in the charge state distribution. At the end of the flare (snapshot 29), the
\ion{Fe}{25} He$\alpha$ feature almost disappears and the Fe L features at $\sim$1~keV
increases, suggesting that Fe$^{24+}$ is taken over by Fe$^{23+}$ (Li-like) and less
charged ions.

\subsection{Spectral modeling} \label{s3-2}
\subsubsection{Phenomenological model} \label{s3-2-1}
We first tried a phenomenological model to explore the characteristics of the data set.
We used the first eight snapshots with sufficient statistics for fitting in the energy
range of interest (5--10~keV). We assessed the level of the persistent emission
using the last part of the datasets (snapshots 24--31 except for 27--29 with a possible rebrightening) when the flare seemed settled. Its contribution to the first part of the flare (snapshots 0–7) is smaller than a fraction of $\sim$ 0.1 in the energy band of interest. We do not use time-sliced spectra
within each snapshot, as we found that they exhibit no significant change between the
former and the latter halves.

The model is composed of two components: a Bremsstrahlung model for the continuum and
Gaussian models for the lines. A total of nine Gaussian lines are added: three for
\ion{Fe}{26} Lyman series (Ly$\alpha$, Ly$\beta$, and Ly$\gamma$), three for
\ion{Fe}{25} $n=2 \rightarrow 1$ series (He$\alpha$, He$\beta$, and He$\gamma$), and
He$\alpha$ lines of other He-like ions of \ion{Cr}{23}, \ion{Mn}{24}, and \ion{Ni}{27}.

The fitting was conducted simultaneously for the two components. The free
parameters for the Bremsstrahlung component are the electron temperature
($T_{\mathrm{e}}$) and the emission measure ($EM$). For the Gaussian components, we
fitted their normalization individually and the energy shift collectively within
$\pm$1\% around the center. The line widths were fixed to null. For the Lyman series,
the energy averaged between ($n$p) $^{2}P_{1/2}$ $\rightarrow$ (1s) $^{2}S_{1/2}$ and
($n$p) $^{2}P_{3/2}$ $\rightarrow$ (1s) $^{2}S_{1/2}$ weighted by their statistical
weights is used as the line center, where $n \in \left\{2,3,4\right\}$. For the He
series, the energy for ($n$p1s) $^{1}P_{1}$ $\rightarrow$ (1s)$^{2}$ $^{1}S_{0}$ is used
among several fine-structure lines that spread over $<$1\% of the line energy for Cr,
Mn, Fe, and Ni. All fits were statistically acceptable. Fig.~\ref{f03-bregau} depicts
the results of the fitting.

\subsubsection{Physical models} \label{s3-2-2}
We next applied three physical models: CIE, ionizing, and recombining plasmas. They are
respectively implemented as \texttt{apec}, \texttt{nei}, and \texttt{rnei} models in the
\texttt{xspec} spectral fitting package \citep{arnaud1996}. The latter two are off-CIE
plasmas in opposite directions. The ionizing plasma model assumes the initial condition
in which ions are neutral and electrons have a temperature higher than the neutral
ions. The degree of Coulomb relaxation is parameterized by the ionization parameter
$\tau = \int_{0}^{t} n_{\mathrm{e}}(t')dt'$, in which $n_{\mathrm{e}}(t')$ is the
time-varying electron density and $t$ is the time from the initial state. The
recombining plasma model assumes the opposite initial condition that ions are ionized
for a given initial temperature and electrons have a lower temperature. Their relaxation
is also parameterized by $\tau$. Recombining plasma conditions might be observed if the cooling via thermal conduction occurs more rapidly than charge state equilibration after impulsive heating of the plasma.

\begin{figure*}[!hbtp]
 \centering
 \includegraphics[width=0.49\textwidth,clip]{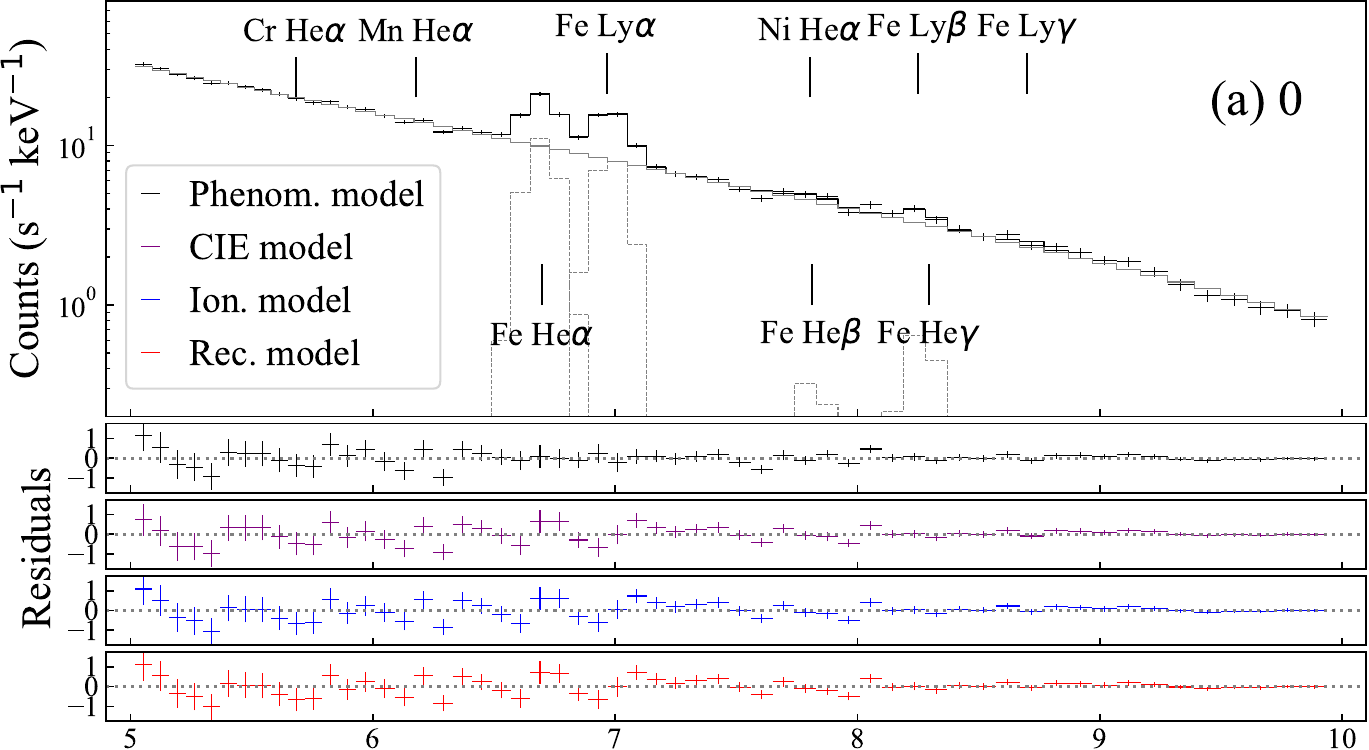}
 \includegraphics[width=0.468\textwidth,clip]{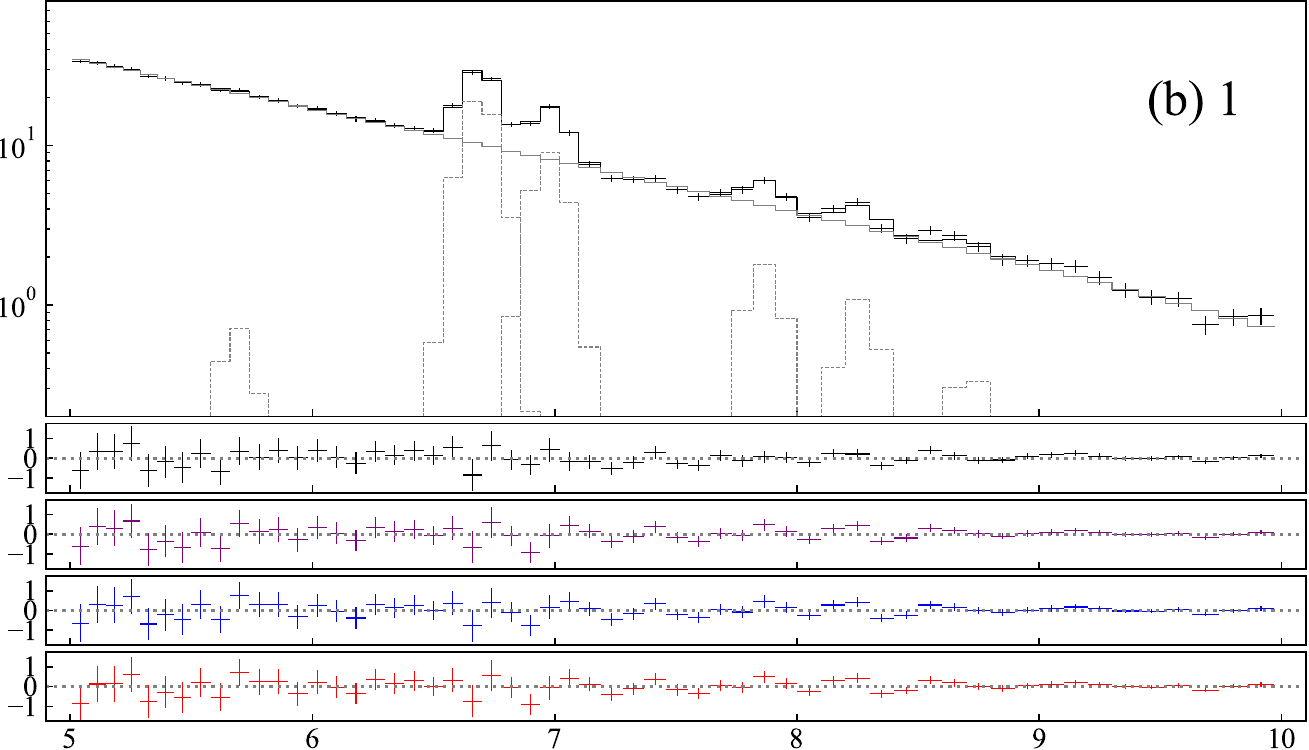}
 \hfill
 \includegraphics[width=0.49\textwidth,clip]{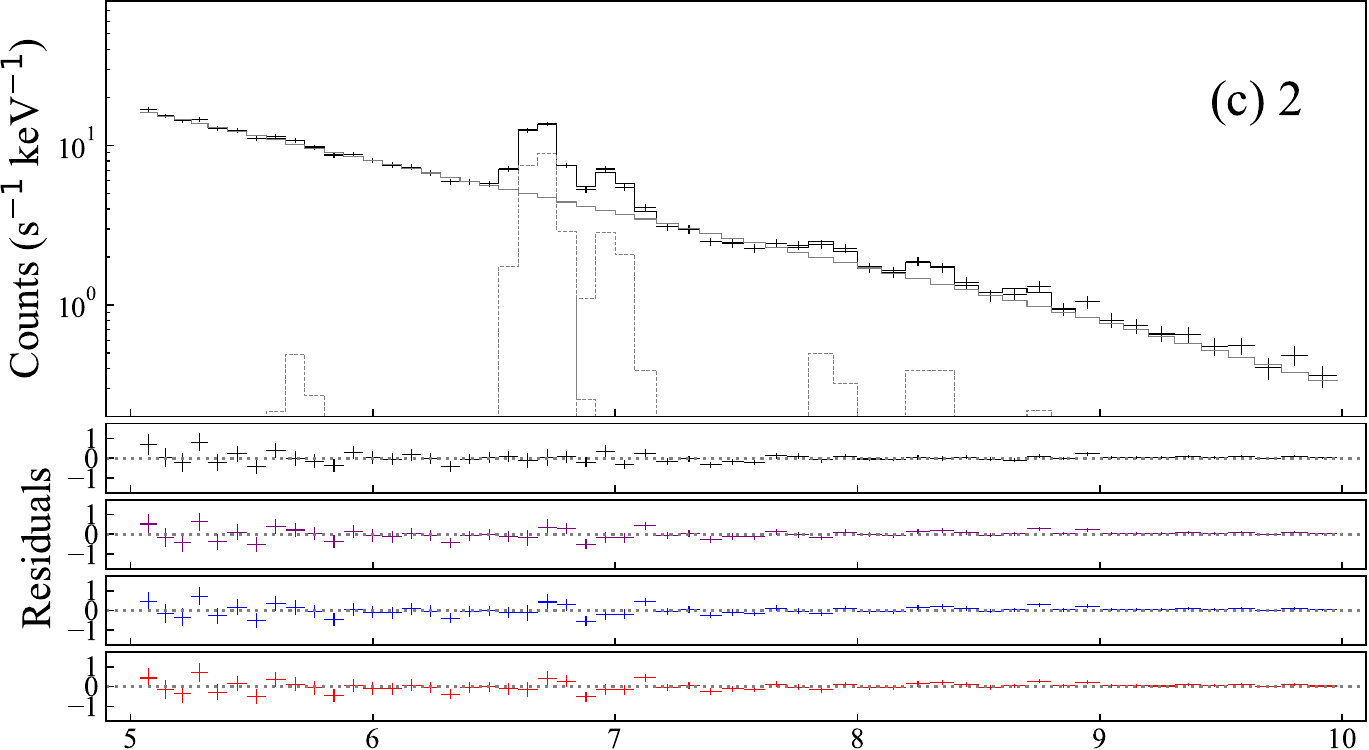}
 \includegraphics[width=0.468\textwidth,clip]{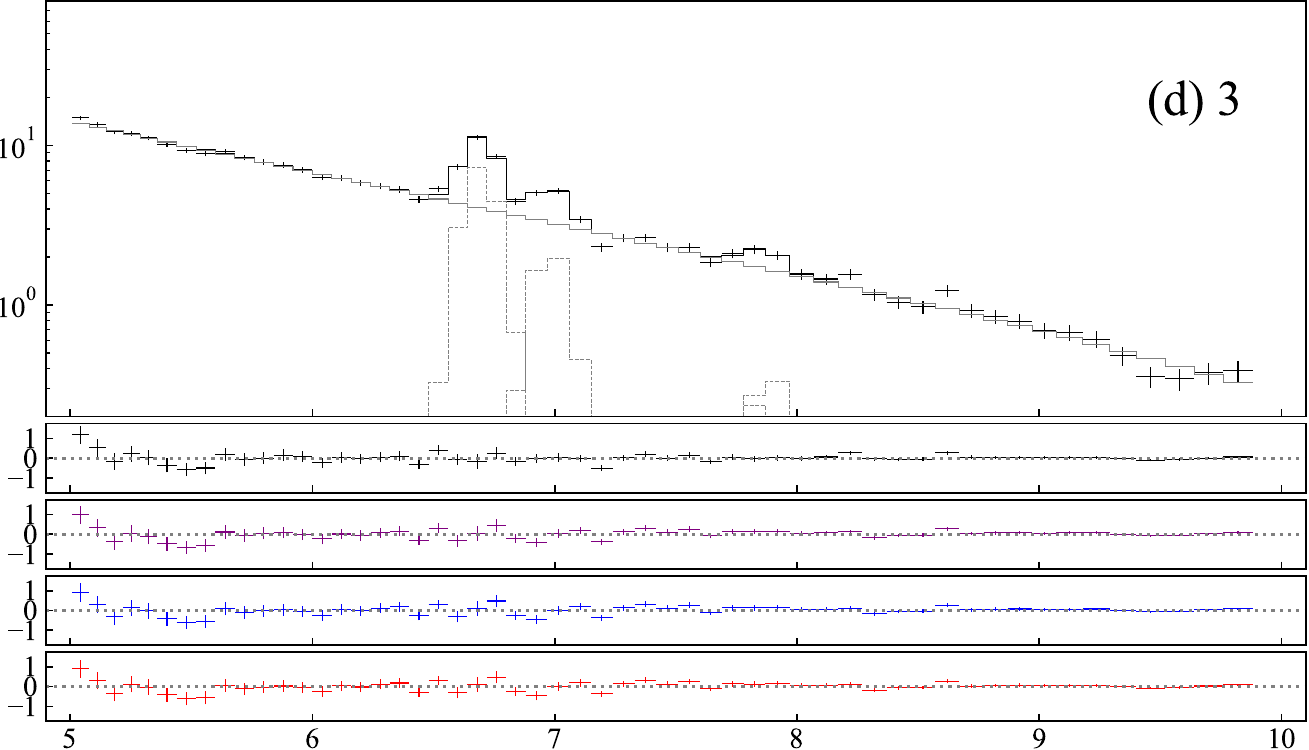}
 \hfill
 \includegraphics[width=0.49\textwidth,clip]{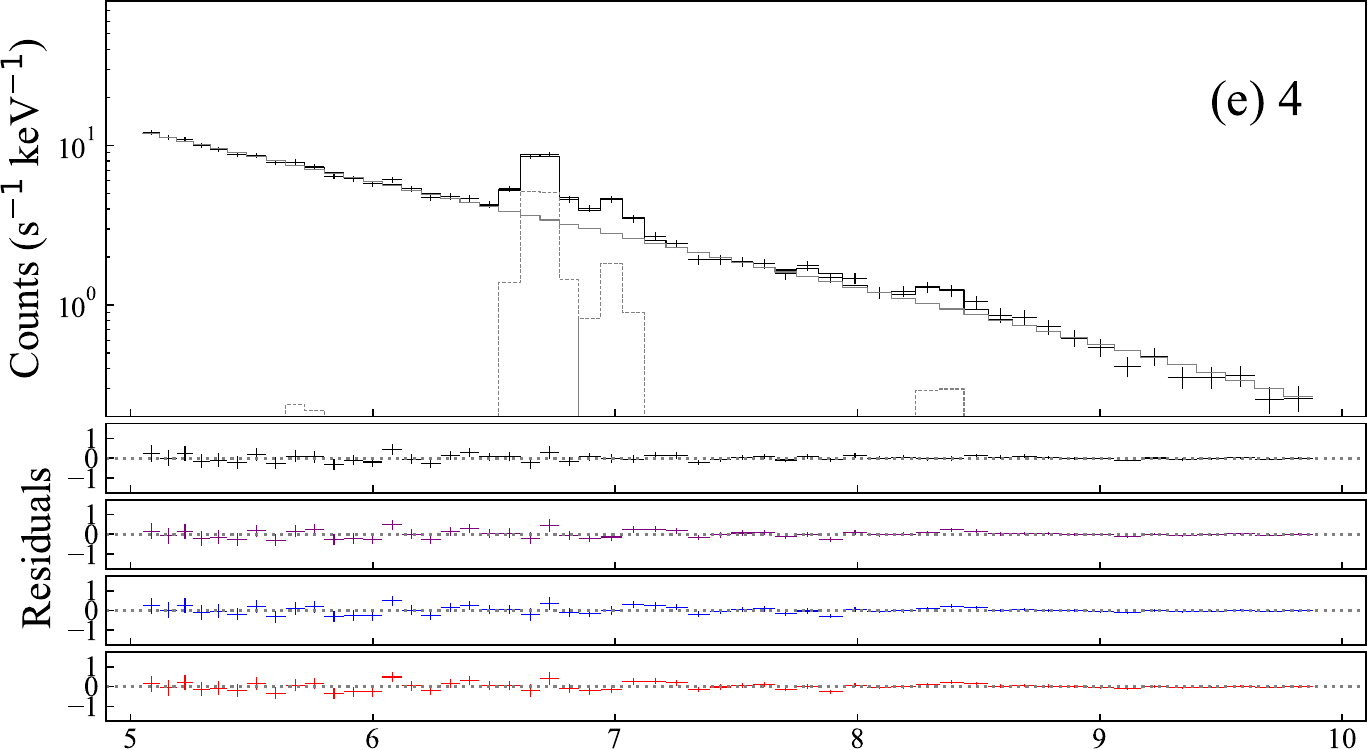}
 \includegraphics[width=0.468\textwidth,clip]{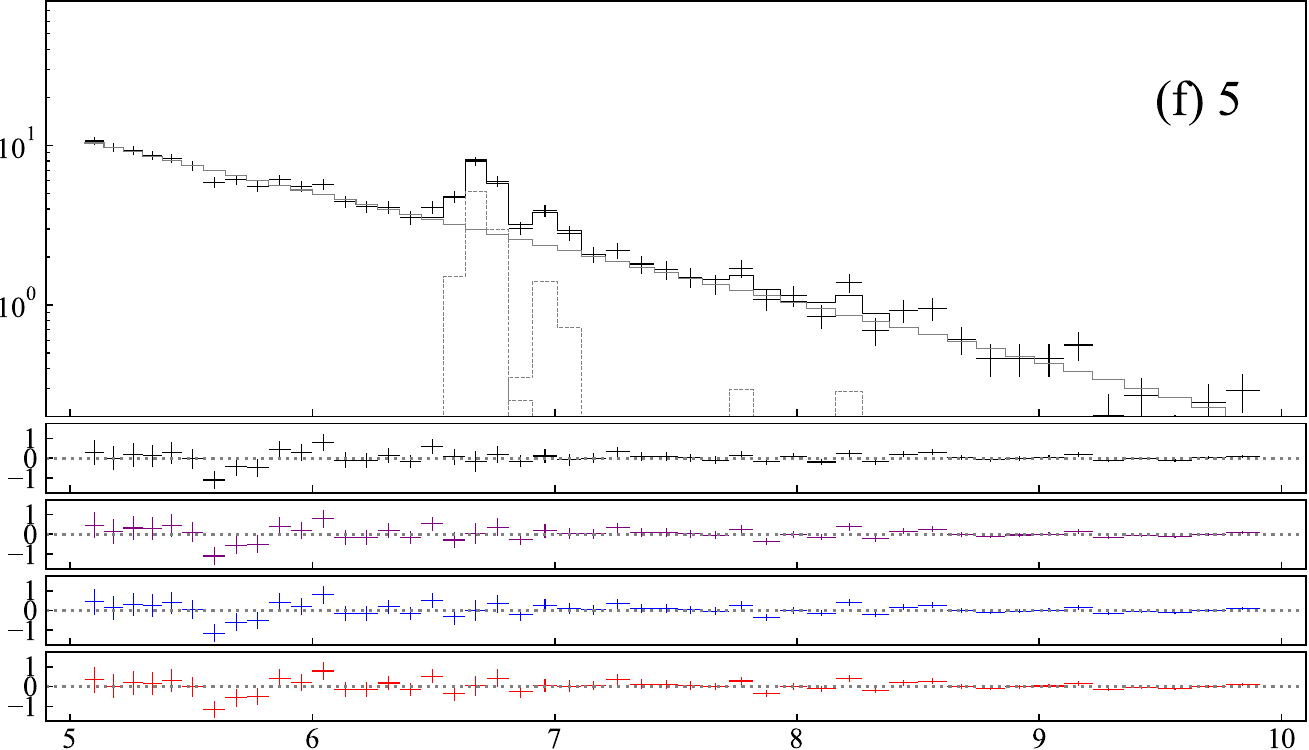}
 \hfill
 \includegraphics[width=0.49\textwidth,clip]{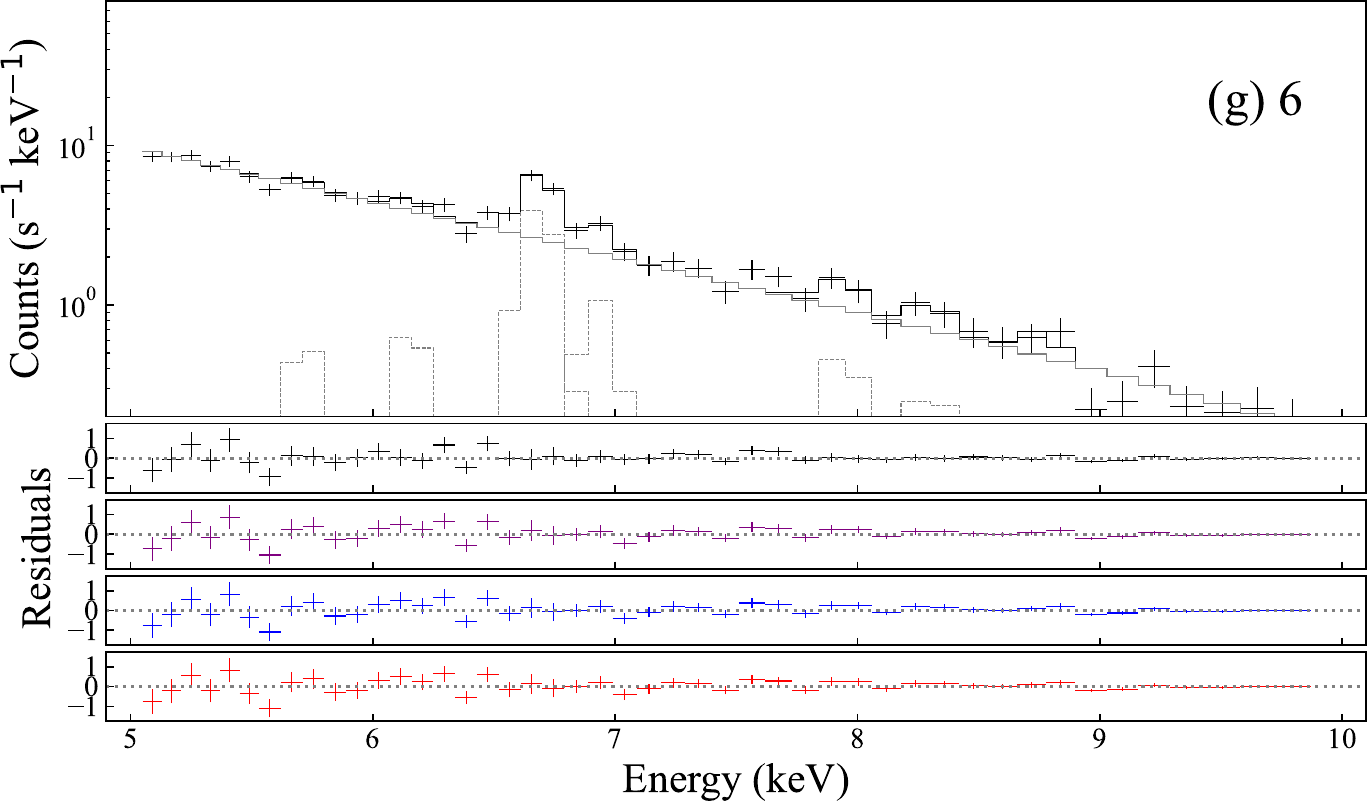}
 \includegraphics[width=0.468\textwidth,clip]{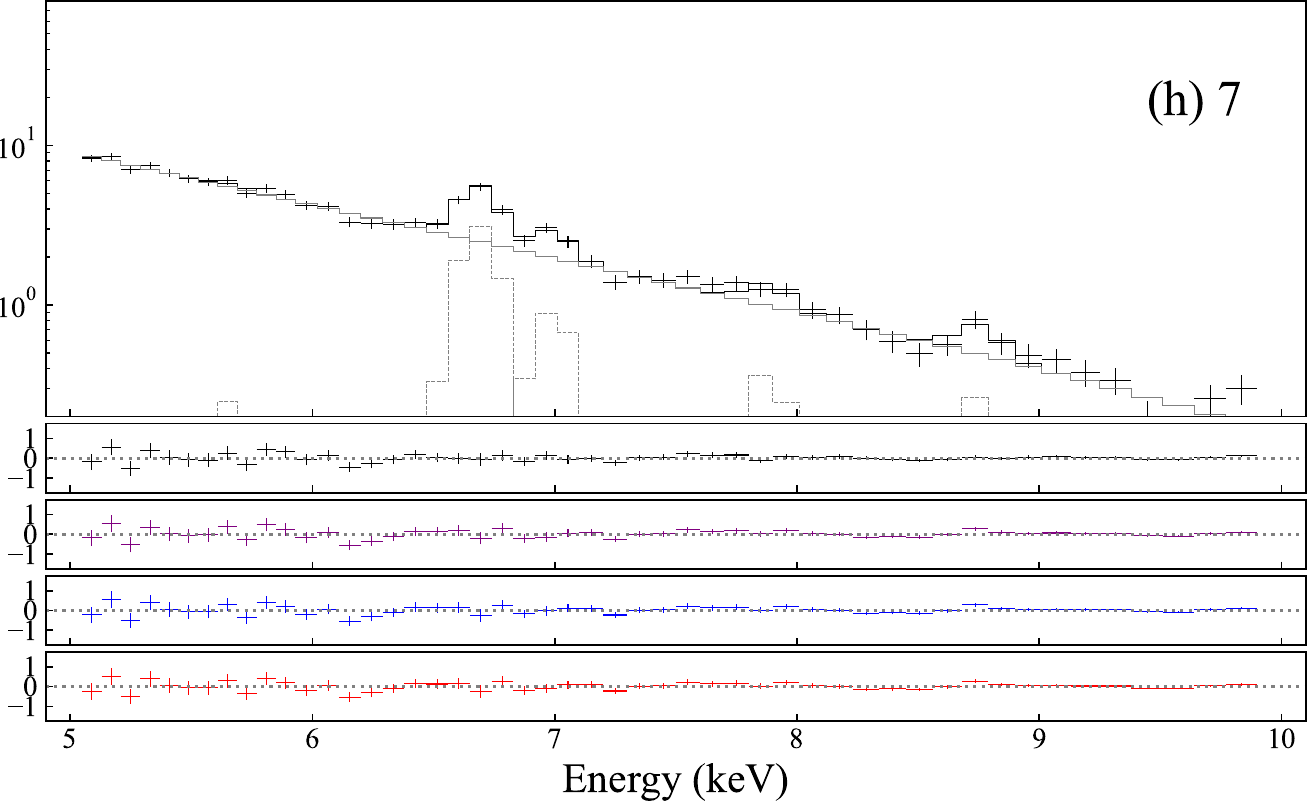}
 \caption{Spectra and best-fit phenomenological model (upper-most panel) and the
 residuals of the fit (other panels) for the phenomenological model (black), CIE plasma
 model (purple), ionizing plasma model (blue), and recombining plasma model (red) for
 snapshots 0--7.}
 \label{f03-bregau} 
\end{figure*}

The free parameters for CIE plasma model are the electron temperature
($T_{\mathrm{e}}$), abundance ($Z_{\odot}$) relative to the cosmic value
\citep{anders1989}, and emission measure ($EM$). For the ionizing and recombining
plasma, the ionization parameter $\tau$ is also added. The initial temperature of the
recombining plasma was fixed to the value obtained in the phenomenological fitting of
snapshot 0. All fits were statistically acceptable. The results of the fitting are
illustrated in Fig.~\ref{f03-bregau}.

\section{Discussion} \label{s4}
\subsection{Continuum observables}\label{s4-1}
\begin{figure}[!hbtp]
\includegraphics[width=1.0\columnwidth,clip]{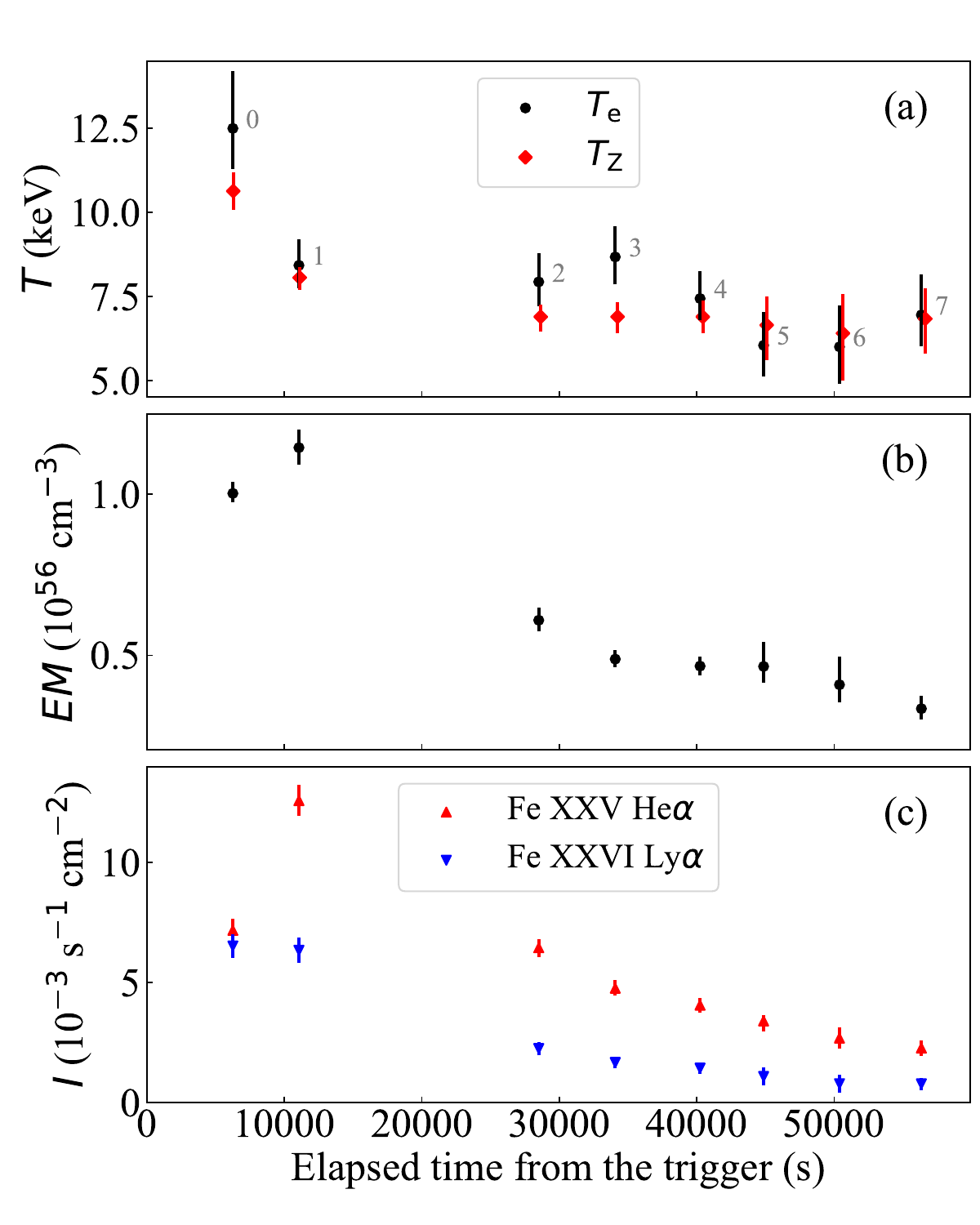}
 \caption{Evolution of best-fit values for $T_{\mathrm{e}}$, $T_{\mathrm{Z}}$,
 $EM$, and line intensities of \ion{Fe}{25} He$\alpha$ and \ion{Fe}{26} Ly$\alpha$ in
 the phenomenological model. The discrepancy between the electron temperature
 $T_{\mathrm{e}}$ and the ionization temperature $T_{\mathrm{Z}}$ suggests an off-CIE
 plasma.}
\label{f04}
\end{figure}

We first explored the continuum observables following the previous work by
\citet{reale2007}. We used the result of the phenomenological spectral fitting
(\S~\ref{s3-2-1}). Fig.~\ref{f04} (a and b) shows the development of the best-fit values
of $T_{\mathrm{e}}$ and $EM$. The peak of $EM$ (snapshot 1) is delayed from that of
$T_{\mathrm{e}}$ (snapshot 0) in time for $\sim 5\times10^3$~s, as in Fig. 1 of
\citet{reale2007}. We assume that the density peak coincides with the $EM$ peak in
snapshot 1. It is notable that $T_{\mathrm{e}}$ and $EM$ do not monotonically
decrease. They stagnate or increase around snapshots 3--4, which may suggest
repeated heating.

Based on the estimated delay time ($\Delta t_{\mathrm{lag}} \sim 5 \times10^3$~s), we can
infer the flare loop half-length $L$ and the peak density $n_{\rm{M}}$ by using
Eqn. (13) and (9) in \citet{reale2007}:
\begin{equation}
   \left(\frac{L}{10^9 ~\mathrm{cm}} \right) = 2.5
    \frac{\psi}{\ln{\psi}}\left(\frac{T_0}{10^7
			   ~\mathrm{K}}\right)^{\frac{1}{2}}\left(\frac{\Delta
			   t_{\mathrm{lag}}}{10^3 ~\mathrm{s}}\right) 
\end{equation}
\begin{equation}
   \left(\frac{n_{\mathrm{M}}}{10^{10} ~\mathrm{cm^{-3}}} \right) =
    13\left(\frac{T_{\mathrm{M}}}{10^7 ~\mathrm{K}}\right)^2\left(\frac{L}{10^9
       ~\mathrm{cm}} \right)^{-1}
\end{equation}
where $\psi = \frac{T_0}{T_{\mathrm{M}}}$ is the ratio of the maximum temperature $T_0$
and temperature at the density peak $T_{\mathrm{M}}$, and $\Delta t_{\mathrm{lag}}$ is
the time lag between $T_{\mathrm{e}}$ and $EM$. These relations are derived from the
assumption that the density reaches its maximum when the radiative cooling rate equals
the conductive cooling rate. $L$ is estimated to be $\sim 3 \times 10^{11}$~cm
($\sim$4~$R_{\odot}$), which is much larger than the typical value for solar flares
($\sim 10^{9}$~cm) and even comparable to the scales of the binary system; the radii of
UX Ari Aa and Ab are 5.6~$R_{\odot}$ and 1.6~$R_{\odot}$, respectively, and the binary
separation is $\sim 0.1$~AU$\sim 20~R_{\odot}$ \citep{hummel2017}. The density maximum
gives a fairly similar value $n_{\mathrm{M}} \sim 4 \times 10^{10}$ cm$^{-3}$ to solar
coronal plasma density in flaring regions \citep{aschwanden1997}. Since the plasma is
confined by the magnetic flux tube, the magnetic field strength ($B$) can be estimated
from the pressure balance. Assuming the plasma to be an ideal gas, $B > \sqrt{16\pi
n_{\mathrm{M}}T_{\mathrm{M}}} \sim 200$~G. By additionally using the measured
$EM$ value, the aspect ratio between the loop length and the loop cross section radius is estimated to
be $\sim$ 0.6. These values are of the same order as those in other giant stellar flares
\citep{favata1999,maggio2000,franciosini2001,reale2004,reale2007,sasaki2021,pillitteri2022}.


\subsection{Line observables}\label{s4-2}
\subsubsection{Phenomenological model}\label{s4-2-1}
We next examined the line observables. Using the phenomenological spectral fitting
(\S~\ref{s3-2-1}), the development of the line intensities of \ion{Fe}{25} He$\alpha$
and \ion{Fe}{26} Ly$\alpha$ is shown in Fig.~\ref{f04} (c). Their intensities are
well-constrained thanks to the photon-rich spectra. In order to examine any deviation
from CIE, the line ratio of Fe Ly$\alpha$ and He$\alpha$ was derived for each snapshot.
The Ly$\alpha$ against He$\alpha$ is larger in snapshot 1 than snapshot 2 despite a
similar $T_{\mathrm{e}}$ value. Here, we used the line ratio of the same elements to
avoid systematics caused by elemental abundance.

\begin{figure}[!hbtp]
 \includegraphics[width=1.0\columnwidth,clip]{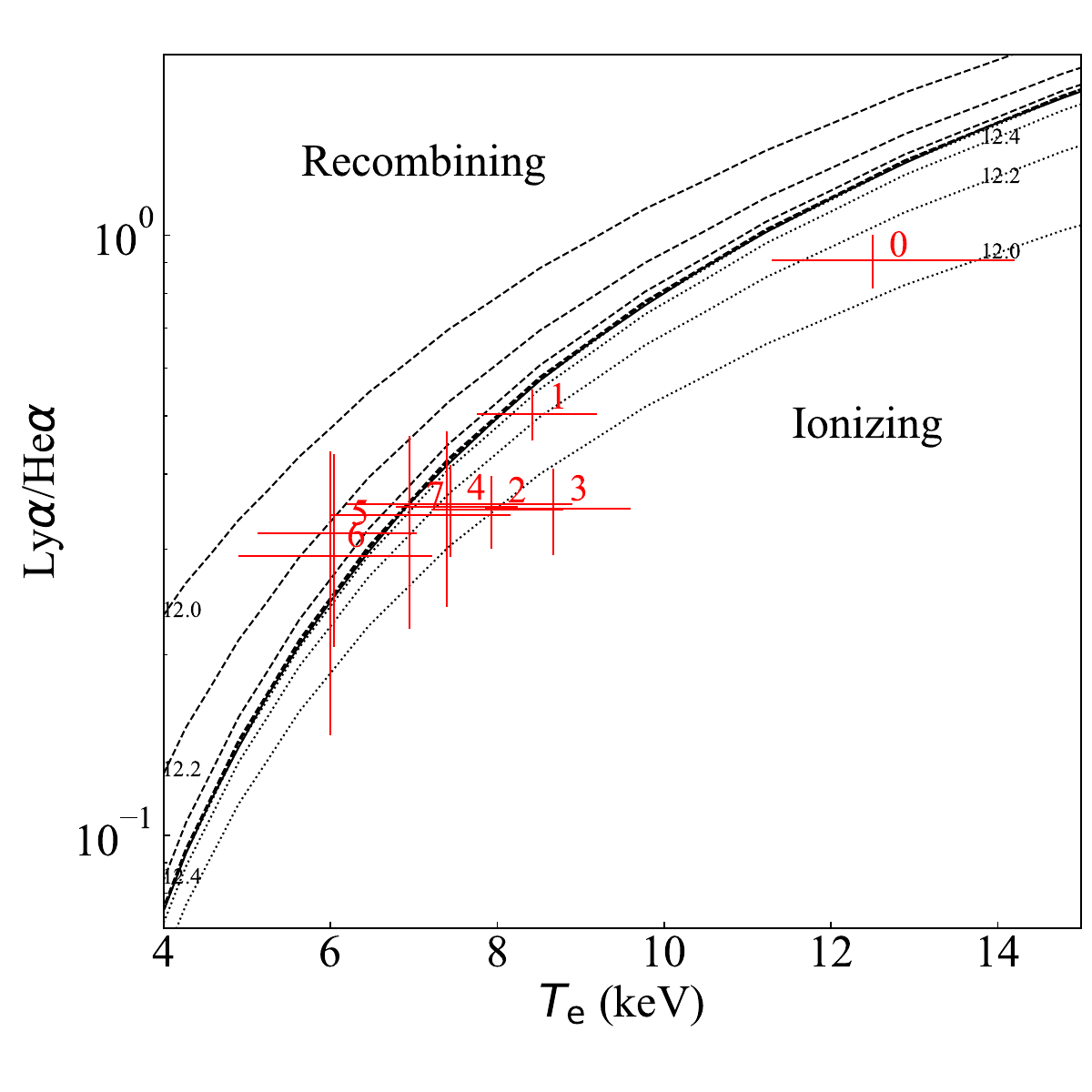}
 \caption{Line ratio of \ion{Fe}{26} Ly$\alpha$ over \ion{Fe}{25} He$\alpha$. Different
 cases are shown in different line styles: CIE (solid), recombining (dashed), and
 ionizing (dotted) plasmas of varying $\log_{10}{\tau}$ (cm$^{-3}$~s) from 12.0 to 13.0
 with a step of 0.2. Data points are plotted with red error bars and the snapshot
 number.}  
\label{f08}
\end{figure}

We calculated the expected line ratio as a function of $T_{\mathrm{e}}$ for CIE,
ionizing, and recombining plasmas (Fig.~\ref{f08}). The starting condition for the
ionizing/recombining plasma is that all Fe ions are neutral/ionized. The line
intensities were extracted from the synthesized X-ray spectra calculated for
$\tau=10^{12}-10^{13}$~cm$^{-3}$~s with geometrical spacing. The Ly$\alpha$ and
He$\alpha$ line complexes that are unresolved with \textit{NICER} were combined.

Based on the calculated results, we plotted the observed results. Snapshot 0
significantly deviates from CIE in the ionizing direction. It approaches CIE for
snapshot 1, but deviates again in the ionizing direction for snapshot 3. The later
snapshots are consistent with CIE within uncertainties. The ionization temperature
($T_{\mathrm{Z}}$) derived from the line ratio assuming CIE is compared with
the electron temperature from the continuum emission ($T_{\mathrm{e}}$) in
Fig.~\ref{f04} (a). The two temperatures are in disagreement in snapshots 1 and 3.

\subsubsection{Physical model}
\begin{figure}[!hbtp]
 \includegraphics[width=1.0\columnwidth,clip]{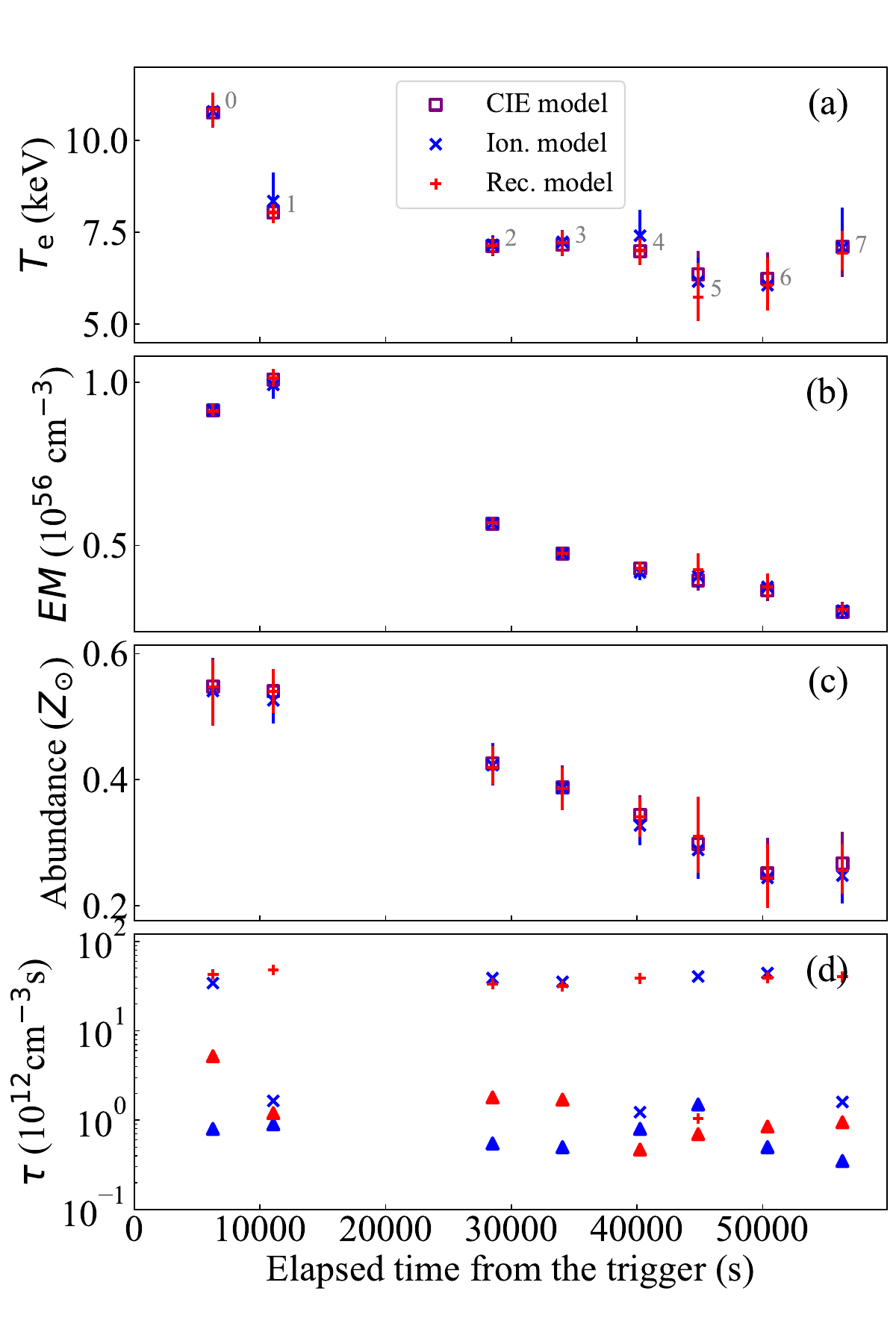}
 \caption{Evolution of best-fit values for $T_{\mathrm{e}}$, $EM$, and abundance of
 the three physical models, and $\tau$ in the physical models. For $\tau$, lower limits
 are illustrated as triangle symbols.}
 \label{nf10} 
\end{figure}

Fig.~\ref{nf10} shows the development of the best-fit values of the parameters of the
three physical models. All the parameters change similarly for the three models,
suggesting that the off-CIE features in the spectra are insignificant.  The best-fit
$\tau$ values for the two off-CIE plasma models are $\gtrsim 1 \times
10^{13}$~cm$^{-3}$~s, which is sufficiently large to reach equilibrium.

The phenomenological model suggests an off-CIE plasma in the ionizing direction in
snapshots 0 and 3 (Fig.~\ref{f08}), whereas the physical ionizing plasma model suggests
CIE condition. This discrepancy arises from the different assessments of
$T_{\mathrm{e}}$. In the former, $T_{\mathrm{e}}=12.5\pm^{1.7}_{1.2}$ and
$8.7\pm^{0.9}_{0.8}$~keV for snapshots 0 and 3. In the latter,
$T_{\mathrm{e}}=10.8\pm^{0.5}_{0.4}$ and $~7.2\pm^{0.3}_{0.4}$~keV. The physical models
determine $T_{\mathrm{e}}$ using not only continuum information but also line
information. Because the energy range of the current dataset does not cover the expected
cut-off of the Bremsstrahlung continuum, the $T_{\mathrm{e}}$ constraint by the
continuum emission alone is loose. As a result, two fitting solutions of high
$T_{\mathrm{e}}$/low $\tau$ or low $T_{\mathrm{e}}$/high $\tau$ are allowed.

\begin{figure*}[!hbtp]
 \centering
 \includegraphics[width=0.49\textwidth,clip]{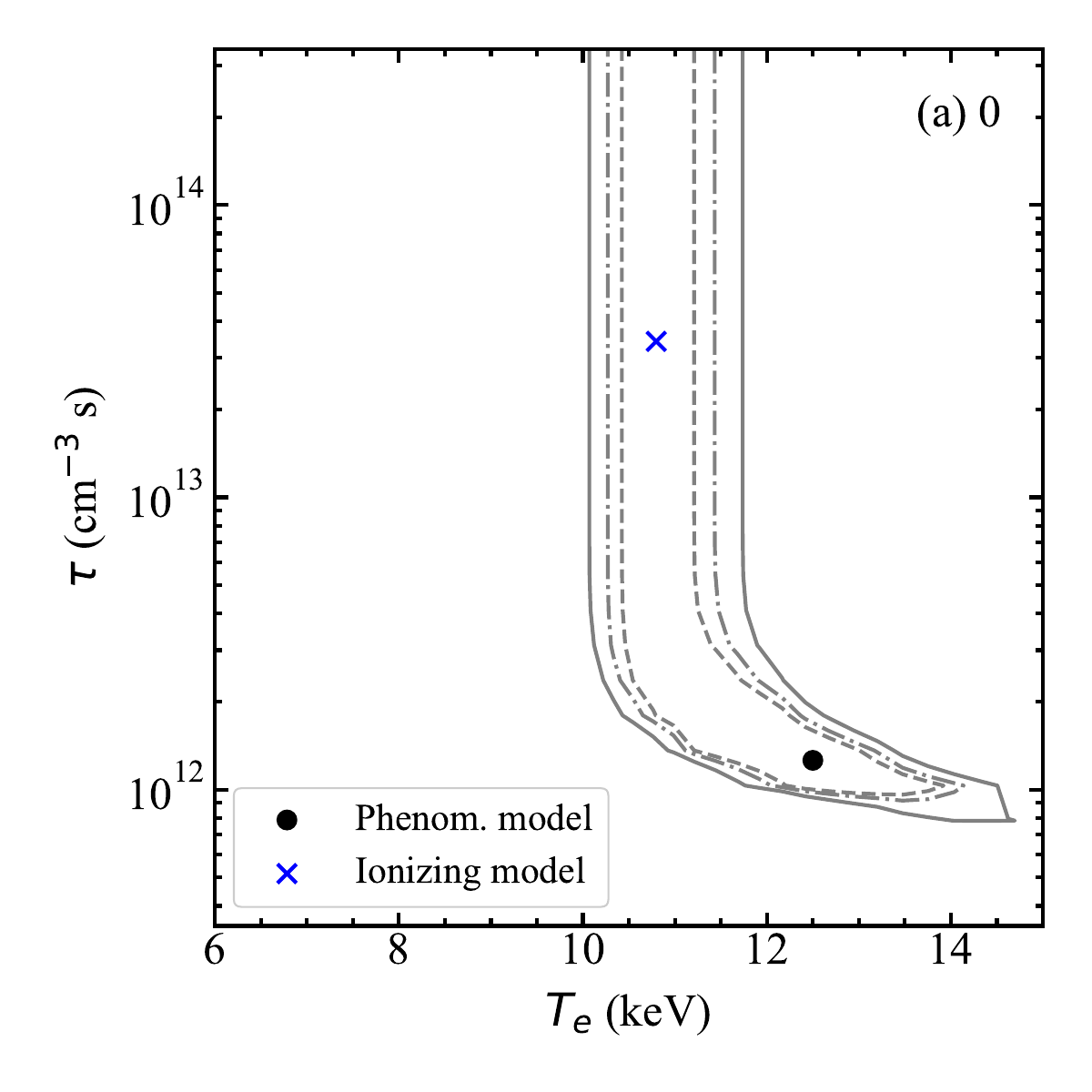}
 \includegraphics[width=0.49\textwidth,clip]{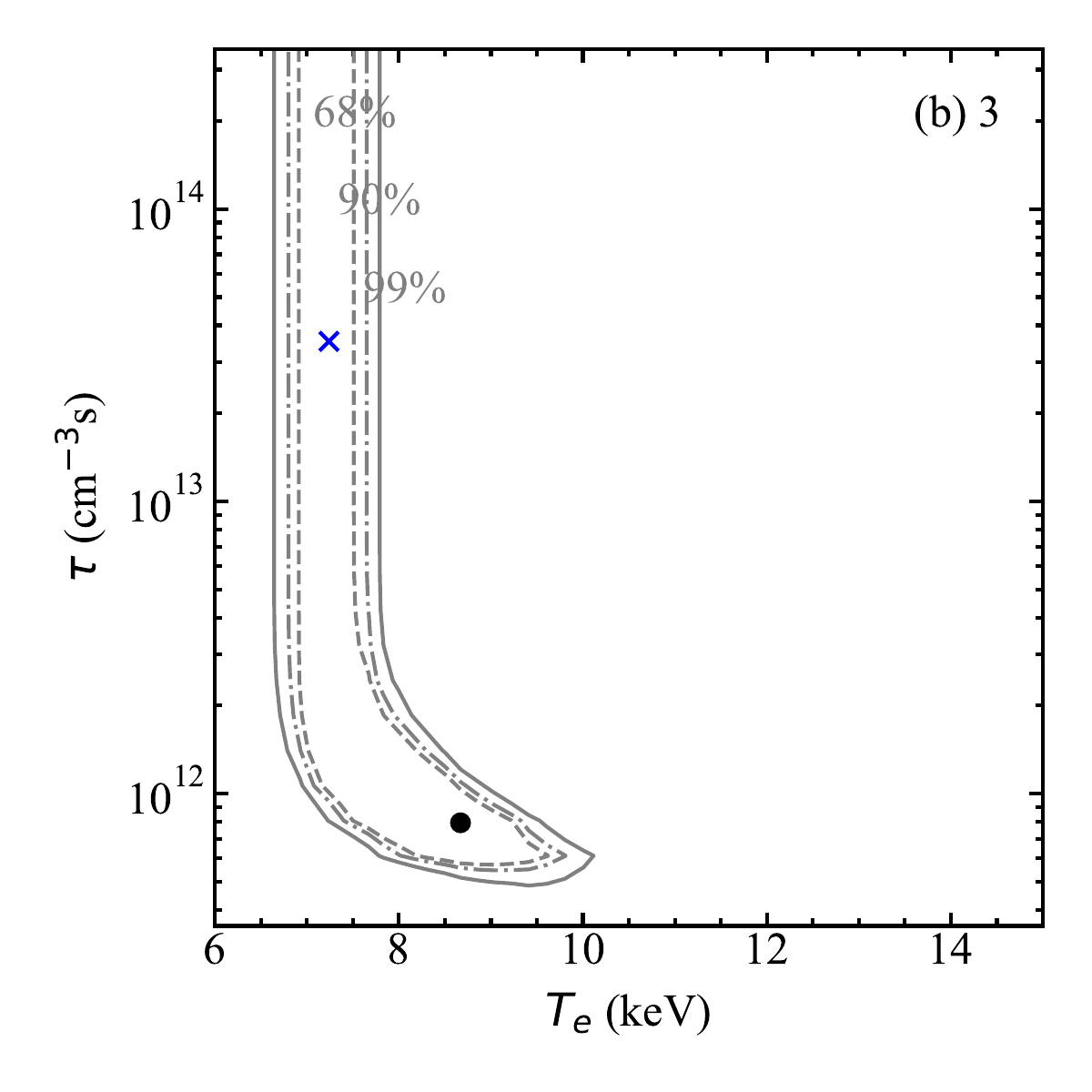}
 \caption{Contour plots of confidence intervals on ($T_{\mathrm{e}}$, $\tau$) for
 snapshots 0 and 3. Dashed, dashed-and-dotted, and solid lines correspond to 68\%, 90\%,
 and 99\%. The best-fit values for the phenomenological model and the physical ionizing
 plasma model are shown with markers. The $\tau$ of the phenomenological model is
 derived from Fig.~\ref{f08}.}
 \label{nf11} 
\end{figure*}

In order to investigate the coupling nature of the two parameters, we made a contour
plot in the ($T_{\mathrm{e}}$, $\tau$) space for snapshots 0 and 3 using the ionizing
plasma model (Fig. \ref{nf11}). The contours indicate that either solution is
acceptable: CIE plasma of $\tau \gtrsim 5 \times 10^{12}$~cm$^{-3}$~s with
$T_{\mathrm{e}}$ more stringently constrained than $\tau$ (thus vertically long part of
the contours) and the ionizing plasma of $\tau \sim 1 \times 10^{12}$~cm$^{-3}$~s with
$\tau$ more stringently constrained than $T_{\mathrm{e}}$ (thus horizontally long part
of the contours). The best-fit parameter pair of the physical ionizing plasma model
resides in the former, whereas that of the phenomenological model resides in the
latter. We conclude that the present spectra do not allow us to distinguish between
these two possibilities.

The development of the light curve favors CIE solution. Assume that there is no
continued heating in the rising phase and the density increases linearly in time as
\begin{equation}
 n(t) = n_0 (t-t_{\mathrm{start}}),
\end{equation}
where $t_{\mathrm{start}}$ is the flare start time that is assumed to be the
\textit{MAXI} trigger time (Fig.~\ref{f01} a). The normalization $n_0$ satisfies $n(t_1)
= n_{\mathrm{M}} = 4 \times 10^{10}$~cm$^{-3}$, in which $t_1$ is the time of the
snapshot 1. Then, the ionization parameter at snapshot 0 ($t_0$) is $\tau_0 =
\int_{t_{\mathrm{start}}}^{t_0} n(t) dt \sim 10^{14}$ cm$^{-3}$~s, which is large enough
to reach an ionization equilibrium.

\section{Conclusion} \label{s5}
Based on the \textit{MAXI} trigger, we observed a giant X-ray flare from UX Ari using
\textit{NICER}. Thanks to the rapid trigger response made possible by the MANGA system,
we successfully captured the rising phase of the flux. Photon-rich spectra were obtained
throughout the flare covered by 32 snapshot observations with \textit{NICER}.

Using the continuum information (temperature, flux, and their peak delays), we
constrained the flare loop size $\sim 3 \times 10^{11}$~cm and the peak electron density
$\sim 4\times10^{10}$~cm$^{-3}$, which are consistent with other giant stellar
flares. Furthermore, using the line information (ratio of the \ion{Fe}{25} He$\alpha$
and \ion{Fe}{26} Ly$\alpha$ lines), we examined any hints of off-CIE plasma. We
fitted the 5--10~keV spectra with a phenomenological model and three physical models
of CIE, ionizing, and recombining plasmas. The X-ray spectra are consistent with CIE
plasma throughout the flare, but the ionizing plasma away from CIE also explains the
spectra in the flux rising phase.

The present study demonstrated the possibility of conducting X-ray spectroscopy studies
of the rising part of stellar flares without relying on luck to capture one during long
exposures. In the near future, we expect that two improvements can be made. One is the
X-ray observation beyond 10~keV using \textit{NuSTAR} \citep{Vievering2019} to constrain
the electron temperature $T_{\mathrm{e}}$ more precisely. The other is
high-resolution spectroscopy using \textit{XRISM}. The microcalorimeter will resolve the
line complex with more constraining power for off-CIE signatures.

\section*{Acknowledgments}
 We appreciate comments by Hiroya Yamaguchi at ISAS. This research used data
 and/or software provided by the High Energy Astrophysics Science Archive Research
 Center (HEASARC), which is a service of the Astrophysics Science Division at NASA/GSFC. This work was supported by JSPS KAKENHI Grant Nos. JP20KK0072 (PI: S. Toriumi), JP21H01124 (PI: T. Yokoyama), JP21H04492 (PI: K. Kusano), 
 JP17K05392 (PI: Y. Tsuboi). 
 Also, Y. T. acknowledges the support by Chuo University Grant for Special Research.

\vspace{5mm}
\facility{
\textit{NICER} \citep{Gendreau2016}, 
\textit{MAXI} \citep{Matsuoka2009}
}
\software{
HEAsoft \citep{2014ascl.soft08004N},
Xspec \citep{arnaud1996},
AtomDB \citep{smith2001,foster2017}
}

\bibliography{main}
\bibliographystyle{aasjournal}
\end{document}